\documentstyle{article}
\input{epsf}
\newcommand{\be}{\begin{equation}}
\newcommand{\ee}{\end{equation}}
\newcommand{\bea}{\begin{eqnarray}}
\newcommand{\eea}{\end{eqnarray}}
\newcommand{\ba}{\begin{array}}
\newcommand{\ea}{\end{array}}

\def\bbox{{\,\lower0.9pt\vbox{\hrule \hbox{\vrule height 0.2 cm
\hskip 0.2 cm \vrule height 0.2 cm}\hrule}\,}}
\newcommand{\dsl}{\pa \kern-0.5em /}

\newcommand{\nn}{\nonumber \\}

\def\bfo{\mbox{\boldmath $\omega$}}
\def\bfO{\mbox{\boldmath $\Omega$}}
\def\bfn{\mbox{\boldmath $\nabla$}}

\newcommand\bomega{\mbox{\boldmath $\omega$}}

\newcommand{\ft}[2]{{\textstyle\frac{#1}{#2}}}
\newcommand{\eqn}[1]{(\ref{#1})}

\def\bfo{\mbox{\boldmath $\omega$}}
\def\bfO{\mbox{\boldmath $\Omega$}}
\def\bfn{\mbox{\boldmath $\nabla$}}

\def\today{\ifcase\month\or
  January\or February\or March\or April\or May\or June\or
  July\or August\or September\or October\or November\or December\fi
 \space\number\day, \number\year}

\input amssym.def
\input amssym.tex
\font\mybb=msbm10 at 10pt
\def\bb#1{\hbox{\mybb#1}}

\def\bR {\bb{R}}

\def\bC {\bb{C}}
\def\bP {\bb{P}}

\def\bfo{\mbox{\boldmath $\omega$}}
\def\bfO{\mbox{\boldmath $\Omega$}}
\def\bfn{\mbox{\boldmath $\nabla$}}

\begin{document}

%%%%%%%%%%%%%%%% title page %%%%%%%%%%%%%%%%%%%%%%%%%%%%%%%%%%%%

\begin{titlepage}
\vfill
\begin{flushright}
QMW-PH-00-07\\
CU-TP-999\\
DAMTP-2000-70\\
hep-th/0012178\\
\end{flushright}

%\centerline{\Large \bf {$\frak{D}\frak{R}\frak{A}\frak{F} \frak {T}$}}
\centerline { \bf \today}

\vfill
\begin{center}
\baselineskip=16pt
{\Large\bf Multi-Domain Walls in Massive Supersymmetric Sigma-Models}
\vskip 0.3cm
{\large {\sl }}
\vskip 10.mm
{\bf ~Jerome P. Gauntlett$^{*,1}$, ~David Tong$^{\sharp,2}$\\
 and  ~Paul K. Townsend$^{\dagger,3}$ } \\
%\\[2mm]
\vskip 1cm
%\vfill
{\small
$^*$
  Department of Physics\\
  Queen Mary, University of London\\
  Mile End Rd, London E1 4NS, UK\\
}
\vspace{6pt}
{\small
 $^\sharp$
Department of Physics\\
Columbia University, \\
New York, NY, 10027, USA\\
}
\vspace{6pt}
{\small
 $^\dagger$
DAMTP,\\
Centre for Mathematical Sciences, \\
Wilberforce Road, \\
Cambridge CB3 0WA, UK\\
}
\end{center}
\vfill
\par

\begin{center}
{\bf ABSTRACT}
\end{center}
\begin{quote}

Massive maximally-supersymmetric sigma models are shown to exhibit 
multiple static kink-domain wall solutions that preserve 1/2
of the supersymmetry. The kink moduli space admits a natural 
K\"ahler metric. We examine in some detail the case when the
target of the sigma model is given by the co-tangent bundle 
of ${\bC \bP}^n$ equipped with the Calabi metric, and 
we show that there 
exist BPS solutions corresponding to $n$ kinks at arbitrary 
separation. We also describe how $1/4$-BPS charged and intersecting domain 
walls are described in the low-energy dynamics on the kink moduli space. 
We comment on the similarity of these results to monopole 
dynamics.

\vfill
 \hrule width 5.cm
\vskip 2.mm
{\small
\noindent $^1$ E-mail: j.p.gauntlett@qmw.ac.uk \\
\noindent $^2$ E-mail: tong@physics.columbia.edu \\
\noindent $^3$ E-mail: p.k.townsend@damtp.cam.ac.uk 
}
\end{quote}
\end{titlepage}
%%%%%%%%%%%%%%%%%%%%%%%%%%%%%%%%%%%%%%%%
\setcounter{equation}{0}
\section{Introduction}

Models admitting vortices, lumps, monopoles or 
instantons typically have BPS limits in which 
the forces between the objects cancel, resulting in 
a moduli space of static multi soliton solutions. The structure
of these moduli spaces carries important kinematical
and dynamical information about the solitons. 
Moreover, they have interesting mathematical properties and appear  
ubiquitously in string theory. It is thus natural to enquire
about the possibility of scalar field theories that might 
exhibit multi-kink solutions with similarly interesting moduli spaces. 

Consider models with BPS kink solutions 
with energy bound $E=|Z|$, where $E$ is the energy
per unit area of the wall and $Z$ is a real central charge
appearing in the supersymmetry algebra. 
The BPS energy bound for two parallel domain walls is 
obviously saturated when they are infinitely separated, so 
reducing the separation cannot decrease the energy. It follows that
the force between the walls at large separation
is either repulsive or zero. This force can be calculated \cite{manton} and 
for models with only a single scalar field it is 
always repulsive. Thus, while there 
may exist  time dependent multi-domain wall 
solutions (such as the kink-anti-kink breather of 
the sine-Gordon model), these models contain no static multi-domain wall 
solutions in which the separation may be chosen arbitrarily.

If we consider kinks carrying a complex, or vectorial, central charge
then two kinks with non-parallel charges may exert an attractive force
on each other, in which case they will eventually fuse into a third
kink carrying a central charge that is the vector sum of the charges
of the initial two kinks. However, it also possible that two kinks
with non-parallel charges will repel each other. Which possibility is
realized depends on the details of the model; 
a Wess-Zumino model in which walls repel or attract according to the
choice of parameters in the superpotential was studied in \cite{pauled}.

The above comments indicate that the simplest models admitting
multi BPS kink solitons should have several
scalar fields.
Multi-kink solutions have been found in 
generalized Wess-Zumino models \cite{Shif1,Shif2}. 
However, these theories have 
four supersymmetries which is not 
sufficient to endow the resulting kink moduli spaces with a great deal of geometric 
structure. The only field theories with {\sl eight} supersymmetries 
that admit static kink solutions are the ``massive'' supersymmetric 
hyper-K\"ahler sigma models, so it is to these models that 
we turn our attention. These typically admit not only
kinks, and their charged counterparts, the Q-kinks \cite{AT2}, 
but also a variety of other BPS solutions, such as Q-lumps 
\cite{abraham}, intersecting domain walls \cite{intersect} 
and D-branes \cite{dbrane}. The purpose of this paper is to exhibit and study 
a class of massive hyper-K\"ahler sigma models that admit {\sl multi-kink} 
(and multi-Q-kink) solutions, for which the moduli space is K\"ahler. 

One might suspect that 
the cancellation of inter-kink forces that is needed for static 
multi-kink solutions to exist is a direct consequence of 
the 8 supersymmetries, but this is certainly not the case.
To see why, consider the sigma model model with a target space metric 
given by the multi-centre ALE 4-metric 
\be
{\rm d}s^2 = U{\rm d}{\bf X}\cdot{\rm d}{\bf X}+U^{-1}({\rm d}\psi+\bomega\cdot
{\rm d}{\bf X})^2
\label{ale}\ee
where $\nabla\times\bomega=\nabla U$. This metric has a 
tri-holomorphic isometry 
associated to the Killing-vector field $\partial_\psi$, and the 
``centres'' of the metric are the isolated fixed points of this vector field. 
The norm of $\partial_\psi$ is, up to a multiplicative factor, the unique
scalar potential term (for this model) that is compatible with all 8 
supersymmetries \cite{agf}. The choice of multiplicative factor 
corresponds to a choice of mass units, so we may take the potential to be 
\be
V=\ft12U^{-1}\, .
\ee
The addition of this potential to the action yields 
a ``massive'' sigma model 
with isolated vacua at the centres of the metric. For  
$N$ colinear centres the harmonic function $U$ is given by,
\be
U=\sum_{i=1}^N \frac{1}{|{\bf X}-m_i{\bf n}|}\, ,
\ee
where ${\bf n}$ is a unit 3-vector and we may order the centres such that 
$m_i<m_{i+1}$. The $N$ vacua are given by ${\bf X}=m_i{\bf n}$, and there 
exist BPS domain walls interpolating between any pair of {\sl adjacent} 
vacua \cite{opf}, each of which preserves (the same) half of 
supersymmetry. However, ${\bf X}\cdot{\bf n}$ is 
the only ``active'' sigma-model field of these solutions, so the calculation 
of the force between two widely separated kinks reduces to a calculation
similar to that of \cite{manton} for models with only a 
single scalar field. This force is non-vanishing. Thus finitely 
separated domain walls interpolating between non-adjacent vacua 
do not exist in this model.

These considerations suggest that one will need to consider
higher-dimensional target spaces to find multi-kink solutions in 
massive hyper-K\"ahler sigma models. Here we consider models for which the target 
space metric is a hyper-K\"ahler Calabi metric on the 
co-tangent bundle $T^\star({\cal N})$, where ${\cal N}$ is a compact 
K\"ahler manifold of complex dimension $n$. If ${\cal N}$ admits a 
holomorphic killing vector then we may construct a supersymmetric massive 
sigma-model on $T^\star({\cal N})$ with 8 supercharges. In fact, the kink 
solutions of this model actually lie within the zero-section of the 
tangent bundle. In other words, they are also solutions to the massive 
sigma-model with 4 supercharges on ${\cal N}$. In the 
following section, we discuss several features 
of kink solutions in these models. The BPS equations describing the spatial 
and temporal variation of a domain wall coincide with the Morse and 
Hamiltonian 
flows of the Killing potential on ${\cal N}$, respectively. 
The domain wall moduli space is therefore identified with the space of Morse flows 
with given fixed points. It is a non-compact manifold 
with a natural K\"ahler metric. 

The simplest Calabi metric has ${\cal N}={\bC \bP}^n$. In the remainder of 
the paper we discuss in detail the domain wall solutions of this model. 
As we shall see in section 3, the 
potential allowed by supersymmetry generically has $(n+1)$ 
isolated vacua, and hence we take $n\ge 2$. 
As with the ALE 4-metrics, these vacua have a natural linear 
ordering. However, in contrast to the ALE case, the domain walls are not 
restricted to lie between between adjacent vacua. Rather, we shall exhibit 
explicit BPS kink solutions interpolating between {\sl each} pair of vacua. 
Moreover, we shall show that the solution which interpolates between the 
$I^{\rm th}$ and $J^{\rm th}$ vacua is part of a moduli space of solutions of 
dimension $2|I-J|$. We show that the 
collective coordinates on this space may be thought of as the position, 
together with an internal degree of freedom, of $|I-J|$ {\sl fundamental} 
domain walls, each of which interpolates between neighbouring vacua. 

In section 4, we discuss the dynamics of domain walls in the $T^\star({\bC\bP}^n)$ 
model. We show that the moduli space metric is toric K\"ahler. We further 
discuss the dynamics of domain walls in the presence of two or more potentials 
and argue that it is given by a massive sigma model on the kink moduli space. 
We explain how this allows one to describe 
$1/4$-BPS Q-kinks and $1/4$-BPS intersecting domain wall solutions in these 
theories.

We end in section 5 with a discussion. We comment on the
similarities of these
results to those for monopole dynamics and mention an application 
to string theory.

\section{Domain Walls and Morse Flows}

Let us first consider a sigma model with $4$ supercharges in 
$D\leq 4$ space-time dimensions with compact 
target space ${\cal N}$ of complex dimension $n$. We endow ${\cal N}$ with 
a K\"ahler metric, $g$, and denote the 
K\"ahler form by $\Omega$ and the complex structure by $J$. In $D\leq 3$ 
dimensions there exists a deformation of this theory, 
consistent with supersymmetry, given by the addition of a potential, 
\be
V=\ft12\mu^2k^2
\label{tatiana}\ee
where $\mu$ is a mass parameter and $k$ is holomorphic 
Killing vector field, which we 
assume to have only isolated, non-degenerate, fixed points.
 
The one-form $i_k\Omega$ (the contraction of $k$ with $\Omega$)
is closed because 
\be
d(i_k\Omega) = (di_k + i_kd)\Omega \equiv {\cal L}_k\Omega = 0\, ,
\ee
where the first equality follows from the closure of $\Omega$ and the 
second equality from the holomorphicity of $k$; it follows that
\be
dH=i_k\Omega\, , 
\label{dh}\ee 
for some locally-defined Killing potential $H$. The integral of $i_k\Omega$ is
a topological charge equal to the difference between the values of $H$ 
at the two endpoints. This topological charge can support a BPS kink, which 
also has a dyonic generalisation known as a Q-kink, carrying  
a Noether charge associated to the Killing vector field $k$. Denoting by 
$\phi^i$ the coordinates on ${\cal N}$, the energy density is given by, 
\be
{\cal E} = {1\over 2}\int {\rm d}x\ g_{ij}
\left(\dot{\phi}^i\dot{\phi}^j+\phi^{i\prime}
\phi^{j\prime}\right)+\mu^2g_{ij}k^{i}k^j \, .
\ee
This may be rewritten as
\bea
{\cal E} &=&\int\!\bigg\{ {\rm d}x\ \ft12 g_{ij}
(\phi^{i\prime}+\mu\cos\alpha J^i_{\ k}k^k)
(\phi^{j\prime}+\mu\cos\alpha J^j_{\ l}k^l)+ 
\mu\cos\alpha\ \frac{\partial H}{\partial x} \nn
&&\ \ \ \ \ \ \ \ +\  
\ft12 g_{ij}(\dot{\phi}^i-\mu\sin\alpha k^i)
(\dot{\phi}^j-\mu\sin\alpha k^j) 
+\mu\sin\alpha\ \dot{\phi}^ik_i\bigg\}
\label{abby}
\eea
for arbitrary angle $\alpha$. Maximising the right hand side 
with respect to $\alpha$, we deduce the 
Bogomol'nyi bound,
\be
{\cal E}\geq\mu\sqrt{T^2+Q^2}
\ee
where 
\be
T=[H]^{+\infty}_{-\infty}\ ,\ \ \ \ \ \ 
Q=\int {\rm d}x\ \dot{X}^ik_i
\ee
are the topological and Noether charges respectively. The 
Bogomol'nyi bound is saturated
for solutions of the equations 
\bea\label{tink}
\dot{\phi}^i&=&\mu\sin\alpha k^i\nn
\phi^{i\prime}&=&-\mu\cos\alpha J^i_{\ k}k^k\, .
\eea
Both of these equations have a natural geometrical meaning. 
Up to rescaling, the temporal evolution of the 
fields is determined by treating $H$ as a Hamiltonian,
\be
\dot{\phi}^i=\frac{\partial H}{\partial \phi^j}\Omega^{ij}
\ee
The spatial evolution arises by treating $H$ as a Morse function on 
${\cal N}$. The non-degeneracy of $\Omega$ ensures that $H$ is a good 
Morse function with critical points at the fixed points of $k$. The 
Morse flow is
\be
\phi^{i\prime} = \frac{\partial H}{\partial \phi^j}g^{ij} 
\label{ebony}\ee
which, again up to rescaling, coincides with the 
spatial Bogomol'nyi equation \cite{witten}. Note that 
the Morse and Hamiltonian flows on ${\cal N}$ are orthogonal. In the 
remainder of this section, we discuss the time independent 
Morse flows in more detail. 

First note that the critical points of $H$ are in one-to-one 
correspondence with the 
vacua of the potential \eqn{tatiana}. At each point some flows will depart  
while others will terminate. The dimension, $p$, of the hypersurface
formed by the Morse flows  
departing from a given critical point is known as the Morse index of that 
point, and it is equal to the number of negative eigenvalues of the covariant 
Hessian $(D^2H/D \phi^i D \phi^j)$ at that point. As we assumed 
the fixed points of $k$ to be non-degenerate, this guarantees 
that the hypersurface formed by the flows 
terminating at a fixed point of Morse index $p$ will have dimension
$(2n-p)$.
Since ${\cal N}$ is K\"ahler, $p$ is even and moreover for the function $H$, 
the usual Morse inequalities are saturated; it follows that the 
number of fixed points with Morse index $p$ is equal to 
the Betti number $B_p$.
In particular, there exists a single critical point with Morse 
index $2n$, from which flows 
only depart, and a single critical point with Morse index $0$, from 
which no flows depart.
  
As a solution to the sigma model with four supercharges and target space 
${\cal N}$, a kink interpolating between a vacuum of index 
$p$ and a vacuum of index $p^\prime$ has $|p-p^\prime|$ fermionic 
zero modes \cite{witten}. Of these only two of 
arise from broken supersymmetries. The remainder 
are ``accidental''. The unbroken supersymmetries then ensure the 
existence of $|p-p^\prime|$ real bosonic zero modes, and hence 
$|p-p^\prime|$ bosonic collective coordinates. The physical 
interpretation of one of these is as the centre-of-mass position
of the domain wall; its complex partner is an angle conjugate 
to the total Noether charge. This pair of collective coordinates
partner the two Nambu-Goldstone fermions arising from 
the two broken supersymmetries. 

What is the physical interpretation of the 
remaining $|p-p'|-2$ bosonic collective coordinates? They could either 
correspond to further internal degrees of freedom or, alternatively, to 
the relative positions and internal coordinates of more than one domain 
wall. Let us see under which circumstances we may expect the latter 
interpretation. Suppose we have three critical 
points with indices $p$, $q$ and $p^\prime$ such that $p>q>p^\prime$. 
Suppose further that there exists a Morse flow $\Gamma$ from 
$p\rightarrow q$ and a second Morse flow $\Gamma^\prime$ from 
$q\rightarrow p^\prime$. Then, by continuity, we expect there 
to exist a flow from $p\rightarrow p^\prime$ which is close to 
$\Gamma\cup\Gamma^\prime$. The speed of this flow, determined by \eqn{ebony}, reduces 
in the vicinity of the critical point $q$, ensuring that the energy density profile of 
the solution looks like two well separated domain walls sandwiching 
the vacuum $q$.

Let $u^a$, $a=1,\cdots,|p-p^\prime|$ be the   
collective coordinates. These 
are promoted to fields of the low-energy effective action for 
the (multi) kink domain wall. This low-energy dynamics is 
again a sigma model but now with a target space metric supplied
by the usual metric on the soliton moduli space, 
\bea
G_{ab}= \int {\rm d}x \ \frac{\partial \phi^i}{\partial u^a}
\frac{\partial \phi^j}{\partial u^b}g_{ij}\, ,
\label{meredith}\eea
which may be thought of as a metric on the space of Morse flows. 
This metric is K\"ahler. To see this we first note that the 
low-energy effective action of the multi-kink domain wall is again a 
supersymmetric sigma model, with the metric (\ref{meredith}) as its 
target space metric. Next, we recall that our sigma model with four 
supersymmetries and target space ${\cal N}$  may be embedded
into a sigma model with 8 supercharges and target space $T^\star({\cal N})$. 
The kink solutions now have $2|p-p'|$ fermionic 
zero modes and preserve four of the eight supersymmetries, 
so the effective kink sigma-model with 
target metric (\ref{meredith}) has four supersymmetries. If we choose 
the maximal spacetime dimension, D=5, for the original massive HK sigma-model 
then we will have an effective D=4 supersymmetric sigma-model governing 
the low energy dynamics of the kink domain walls in this D=5 spacetime.
The target space of such a sigma model is necessarily K\"ahler. 

\section{Domain Walls in $T^\star({\bC\bP}^n)$}

In this section we discuss in detail the domain walls for ${\cal N}={\bC\bP}^n$, 
working with the toric HK $4n$-metric on 
$T^\star({\bC\bP}^n)$ with coordinates 
$({\bf X}^I,\psi^I)$ ($I=1,\cdots,n$). 
The Calabi metric is,
\be
ds^2=U_{IJ}d{\bf X}^I\cdot d{\bf X}^J+(U^{-1})^{IJ}(d\psi_I+A_I)(d\psi_J+A_J)
\label{metric}\ee
where,
\be
A_I= d{\bf X}^J\cdot \bfo_{JI}\, ,\qquad 
\bfn_{(J} \times \bfo_{K)I} = \bfn_J U_{KI}\, .
\ee
The functions $U$ are given by
\be
U_{IJ}=\frac{\delta_{IJ}}{X^I}+\frac{1}{|{\bf m}-\sum_{K=1}^N{\bf X}^K|}
\ee
where ${\bf m}$ is a constant 3-vector and the lack of $I,J$ indices 
in the second term implies that it appears in each component of the matrix. 
The triplet of K\"ahler forms are
\be
{\bfO}=({\rm d}\psi_I+A_I){\rm d}{\bf X}^I-\ft12U_{IJ}{\rm d}{\bf X}^I\times
{\rm d}{\bf X}^J
\ee
where the wedge product of forms is implicit. This metric 
appears in physics as the moduli space of a single $U(n+1)$ instanton on  
non-commutative ${\bR}^4$, where the 3-vector ${\bf m}$ is 
related to the the (anti-self-dual) non-commutativity parameter (see, for 
example, \cite{leeyi}). In particular, the $n=1$ Calabi 4-metric coincides with the 
Eguchi-Hanson metric on the $N=2$ ALE space \eqn{ale}. The $4n$-metric has 
$SU(n+1)$ tri-holomorphic 
isometry. In the above coordinates only the Cartan sub-algebra is manifest
corresponding to the 
Killing vector fields $k^I=\partial/\partial \psi_I$. These permit 
the construction of a potential compatible with supersymmetry 
given by the square of the length of a linear combination of these vectors 
\cite{agf}, say $\mu_Ik^I$ for constant $\mu_I$,
\be
V=\ft12\mu_I\mu_J(U^{-1})^{IJ}
\label{pot}\ee
In fact, as shown in \cite{bak,intersect}, this is not the most general 
potential allowed by supersymmetry. For theories with $D\leq 6$ space-time 
dimensions one may sum the squares of the lengths of $(6-D)$ independent, 
mutually commuting, tri-holomorphic Killing vectors. In the following section 
we will consider this possibility, but for now we restrict ourselves 
to the simplest potential given in equation \eqn{pot}.

It will prove useful to define a $(n+1)^{\rm th}$ coordinate,
\be
{\bf X}^{n+1}={\bf m}-\sum_{I=1}^n{\bf X}^I
\label{n}\ee
so that $\sum_{I=1}^{n+1}{\bf X}^I={\bf m}$. The potential \eqn{pot} is given 
explicitly by,
\bea
V=\ft12\sum_{I=1}^N(\mu_I^2X^I)-
\ft12\frac{\left(\sum_{I=1}^N\mu_IX^I\right)^2}{\sum_{J=1}^{n+1}X^J}
\label{potty}\eea
Note that the denominator of the second term is not given by 
$\sum_{J=1}^{n+1}X^J=m$ unless ${\bf X}^I\cdot{\bf m}\geq 0$ for each $I$. In 
fact, this constraint on the coordinates is precisely the restriction to the ${\bC\bP}^n$ 
base of the manifold. We shall not impose this constraint for now, although 
we shall later see that all BPS kinks do in fact lie within this submanifold. 

For generic choice of constants, $\mu_I\neq\mu_J$, 
the potential \eqn{potty} has 
$n+1$ isolated vacua, given by
\be
{\bf X}^I={\bf m}\,\delta^{IJ}\ \ \ \ \ \ \ \ \ \mbox{for $J=1,\cdots,n+1$}
\label{vacua}\ee
For non-generic potentials there is an enlarged moduli space of vacua. 
Specifically, 
if $l$ of the constants $\mu_I$ coincide, then there is a $3(l-1)$ dimensional 
sub-manifold of the Calabi metric with vanishing potential. 
We will consider only generic potentials and examine the kinks that 
interpolate between the different isolated vacua. The relevant Morse function 
is $H=\sum_{I=1}^n\mu_I{\bf X}^I\cdot{\bf n}$. Setting all time 
derivatives to zero, the Bogomol'nyi equations are
\bea
{\bf X}^{I\prime}&=&(U^{-1})^{IJ}\mu_J{\bf n} \label{boggy}\\
\psi^{I\prime}&=&\bomega_{IJ}\cdot{\bf X}^{J\prime}\label{boggy2}
\eea
where the unit 3-vector ${\bf n}=\pm{\bf m}/m$ depending on whether we are 
considering a kink or anti-kink. A BPS kink interpolating between the 
$I^{\rm th}$ and $J^{\rm th}$ vacua, with $I,J=1,\cdots, n$ has energy,
\be
E_{IJ} = m|\mu_I-\mu_J|
\ee
while a kink which interpolates between the $I^{\rm th}$ vacuum 
and the $(n+1)^{\rm th}$ vacuum has mass,
\be
E_{I,n+1} = m|\mu_I|
\ee
We may write these formulae in a unified form if we introduce the $(n+1)$ quantities 
$\nu_I$ such that 
\be
\mu_I=\nu_I-\nu_{n+1}\ \ \ \ \ \ (I=1\cdots,n)
\label{sue}\ee
and the mass of a kink interpolating between the $I^{\rm th}$ 
and $J^{\rm th}$ vacua is now given by
\be
E_{IJ}=m|\nu_I-\nu_J|
\label{hree}\ee
Importantly, rewriting the energy in this fashion also makes it clear that there 
is an ordering to the vacua given by the relative values of $\nu_I$, allowing us 
to talk of neighbouring, or adjacent, vacua. We choose the ordering $\nu_I>\nu_{I+1}$. 
Notice that the form of the energy \eqn{hree} is already suggestive of the 
existence of multi-kink solutions since, assuming $J<I$, we may write,
\be
E_{IJ}=\sum_{K=J}^{I-1} E_{K+1,K}
\ee
Taken at face value, this suggests that the kink may be decomposed into 
$d=(I-J)$ kinks, each of which interpolates between neighbouring vacua. We will 
refer to the kink that interpolates between the $I^{\rm th}$ and $(I+1)^{\rm th}$ 
vacua as the $I^{\rm th}$ {\sl fundamental} kink. An analysis of the supersymmetry 
transformations \cite{intersect} reveals that each of these fundamental kinks preserves 
the same half of supersymmetry, as would be expected if multi-kink solutions 
were to exist. However, one must be wary in drawing such conclusions from the 
Bogomol'nyi energy bound alone. Indeed, all the above statements apply equally 
well to kinks in the ALE 4-metrics discussed in the introduction but, as we
noted there, in this case there simply do not exist BPS domain wall solutions 
interpolating between non-adjacent vacua. Nevertheless, we shall see that in 
the present case the above conclusions are in fact correct. 

We start our analysis of the Bogomol'nyi equations by presenting explicit 
solutions between {\sl any} pair of vacua with $J<I$, 
\bea
{\bf X}^K&\rightarrow&{\bf m}\,\delta^{KI}\ \ \ \mbox{as $x\rightarrow-\infty$} \nn
{\bf X}^K&\rightarrow&{\bf m}\,\delta^{KJ}\ \ \ \mbox{as $x\rightarrow+\infty$}
\label{bound}\eea
with $I,J=1,\cdots,n+1$. We make the ansatz ${\bf X}^K=0$ for $K\neq I,J$ which, 
given the constraint \eqn{n}, requires that the two remaining coordinates sum to 
${\bf X}^I+{\bf X}^J={\bf m}$. Geometrically, this restricts us to a sub-manifold 
$T^\star({\bC\bP}^1)$ where the $n(n+1)$ choices of vacuum pairs reflect the 
$n(n+1)$ natural embeddings of ${\bC\bP}^1$ in ${\bC\bP}^n$. The Bogomol'nyi 
equations now reduce to those on the Eguchi-Hanson space whose solutions were 
given in \cite{AT2},
\bea
{\bf X}^I&=&\ft12{\bf m}-\ft12{\bf m}\ {\rm tanh}\left(\ft12(\nu_J-\nu_J)
(x-x_0)\right) \nn
{\bf X}^J&=&\ft12{\bf m}+\ft12{\bf m}\ {\rm tanh}\left(\ft12(\nu_J-\nu_I)
(x-x_0)\right)
\label{solution}\eea
Given these solutions, the second Bogomol'nyi equation may be solved by simply 
choosing a gauge in which $\bomega$ vanishes over the trajectory \cite{paptown} 
and setting 
$\psi_I=-\psi_J=\varphi_0$ to constant. Thus this kink solution has 
$2$ collective coordinates given by the position, $x_0$, and the 
internal degree of freedom $\varphi_0$. We shall now show that the complete 
solution involves a further $2(d-1) = 2(I-J-1)$ collective coordinates, 
corresponding to the possibility of separating the domain wall \eqn{solution} 
into $d$ fundamental kinks. 

Firstly we prove that for the domain wall with boundary conditions \eqn{bound}, 
any solution to the BPS equations necessarily has ${\bf X}^K\equiv 0$ for 
all $K<J$ and for all $K>I$. To see this, note firstly that the Bogomol'nyi 
equations require ${\bf X}^K\propto {\bf n}$ for all $K$, and so take the form,
\be
{\bf X}^{K\prime}\cdot{\bf n}=\left(\nu_K-\frac{\sum_{L=1}^{n+1}\nu_LX^L}{\sum_{M=1}^{n+1}X^M}
\right)X^K
\label{bog}\ee
Moreover, unlike \eqn{boggy}, this form is also valid for the $(n+1)^{\rm th}$ 
coordinate \eqn{bound}. Near the two end-points \eqn{n} of the domain wall 
trajectory, these equations approximate to 
\bea
{\bf X}^{K\prime}\cdot{\bf n} &\approx & (\nu_K-\nu_I)X^K  
\ \ \ \ \mbox{as $x\rightarrow -\infty$}\nn
{\bf X}^{K\prime}\cdot{\bf n} &\approx & (\nu_K-\nu_J)X^K 
\ \ \ \ \mbox{as $x\rightarrow +\infty$}
\eea
Thus we see that for $K<J$ and for $K>I$, the functions $X^K$ must either vanish 
or have at least two stationary points. Similarly, for $J<K<I$, the functions 
must have at least one stationary point while for $K=I$ and $K=J$, they may be 
monotonic. However, from \eqn{bog}, we see that $X^K$ is stationary at $X^K\neq 0$ 
only if,
\bea
\sum_{L=1}^N(\nu_K-\nu_L)X^L=0
\eea
If we first examine $L=1$ (and assume that $J\neq 1$) then 
$(\nu_1-\nu_L)<0$ for all $L$ and there are no non-trivial solutions to 
the stationary point equation. Thus $X^1\equiv 0$. By induction, the same 
is true for all $X^K$ with $K<J$ and $K>I$. Similarly, this analysis 
implies that $X^I$ and $X^J$ have no stationary points and are 
therefore monotonic. However, it does not rule out the possibility of 
stationary points for $X^K$ with $J<K<I$. 

The above result allows us to restrict attention to domain walls interpolating 
between the first and last vacua (i.e. with 
boundary conditions $J=1$ and $I=n+1$). We will now examine the Bogomol'nyi equations 
inductively, starting with the simplest model admitting 
multi-kink solutions: $T^\star({\bC\bP}^2)$.

\subsection*{$n=2$}

In the previous section we worked with an over-complete set of variables 
subject to the constraint \eqn{n} in order to elucidate the vacuum structure 
of the theory. In this subsection, we revert to the original coordinates 
\eqn{metric}. The ordering of the vacua described in the previous subsection 
is equivalent to the choosing the potential $\mu_1>\mu_2>0$. The BPS equations 
for ${\bf X}^K$ are,
\bea
{\bf X}^{1\prime}\cdot{\bf n}&=&\left(\mu_1-\frac{\mu_1}{m}X^1-\frac{\mu_2}{m}X^2\right)X^1 \nn
{\bf X}^{2\prime}\cdot{\bf n}&=&\left(\mu_2-\frac{\mu_2}{m}X^2-\frac{\mu_1}{m}X^1\right)X^2
\eea
The fixed points of these equations are the vacua \eqn{vacua} of the theory. 
There are three such points,
\bea
{\rm vacuum\ 1}&:&X^1=m\ ,\ X^2=0 \nn
{\rm vacuum\ 2}&:&X^1=0\ ,\ X^2=m \nn
{\rm vacuum\ 3}&:&X^1=0\ ,\ X^2=0 
\label{n2bog}\eea
These lie at the three corners of a right-angle isoceles triangle, with the 
right-angle at fixed point 3. The three 
BPS kink solutions given in equation \eqn{solution} form the sides of this 
triangle, with a fixed direction. Specifically, the kinks interpolate between the 
vacua  $2\rightarrow 1$, $3\rightarrow 2$ and $3\rightarrow 1$. 

Near the fixed point 3, the trajectories are
\be
(X^1,X^2)\approx(e^{\mu_1x},e^{\mu_2x})
\ee
so, for positive $\mu_I$, all trajectories start with a straight line through 
the origin into the triangle. Trajectories can only end at fixed points or 
at infinity. Moreover, they may not cross. Therefore, all those that enter 
the triangle must end on fixed points. The only one that may end on fixed 
point 2 is the $X^1=0$ kink. All others must end on fixed point 3, so there 
is a one-parameter family of trajectories that begin at fixed point 3 and end on 
fixed point 1\footnote{This was also noted by Kimyeong Lee and Piljin Yi 
\cite{kimpil}.}. This is sketched in figure 1. 

\begin{figure}
\begin{center}
\epsfxsize=2.0in\leavevmode\epsfbox{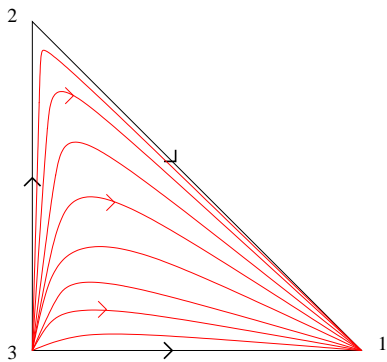}
\end{center}
\caption{\em The BPS flows in the ${\bC\bP}^2$ massive sigma-model. There 
exists a one-parameter family of kink trajectories corresponding to the 
separation of two kinks. The 
two trajectories $3\rightarrow 2$ and $2\rightarrow 1$ may be thought of as 
the limit of 
infinitely separated kinks. 
The straight-line trajectory $3\rightarrow 1$ corresponds to the two kinks with 
zero separation.}
\label{fig1}
\end{figure}

Note that the asymmetry between points 1 and 2 arose from the choice $\mu_1>\mu_2$. 
In one limit of this parameter we have the straight-line $3\rightarrow 1$ kink of 
equation \eqn{solution}. In the other limit, we approach arbitrarily close to 
the union of the trajectories of the $3\rightarrow 2$ kink and the 
$2\rightarrow 1$ kink. This limit 
itself corresponds to the $3\rightarrow 2$ and $2\rightarrow 1$ kinks at 
infinite separation, but at 
any point short of this limit the kinks have finite separation. As the 
separation is decreased, the two kinks eventually merge to form the single 
$3\rightarrow 1$ kink. 
It is natural to call the $3\rightarrow 2$ and
$2\rightarrow 1$ kinks ``fundamental'' kinks, and the family of 
$3\rightarrow 1$ kink solutions as a moduli space of multi-kink solutions.

The fundamental kinks have a single real 
relative collective coordinate. 
Supersymmetry requires that this is paired with a complex partner, such that 
the relative moduli space is K\"ahler. This additional 
collective coordinate comes 
from the angular coordinates $\psi_I$, satisfying equation \eqn{boggy2}.
The multi-kink solutions thus have a four-dimensional moduli space of
solutions.

All of the kink trajectories lie within the triangle depicted in figure 1, 
ensuring that they may not escape to infinity in field space. This triangle 
is the toric diagram for ${\bC\bP}^2$, the zero-section of the Calabi bundle 
(see for example \cite{leung}). The two periodic variables $\psi_I$ provide 
a torus ${\bf T}^2$ which fibred over the triangle to reconstruct ${\bC\bP}^2$. 
Thus, the kinks described above are equally solutions to the ${\bC\bP}^2$ 
sigma-model. We will return to this point in section 4.

On each trajectory, there is a unique value of $Y$ for each value of $X$. 
the trajectories can therefore be described by some curve $Y(X)$. To find 
these curves, we divide the Bogomol'nyi equations \eqn{n2bog}, to get
\be
\mu_1\left({\rm d}X^1+\frac{m-X^1}{X^2}{\rm d}X^2\right)-\mu_2\left({\rm d}X^2+
\frac{m-X^2}{X^1}{\rm d}X^1\right)
=0
\ee
Multiplying by the integrating factor $(m-X-Y)^{-1}$, we deduce that
\be
{\rm d}\log\left(X^{\mu_2}Y^{-\mu_1}(m-X-Y)^{\mu_1-\mu_2}\right)
\ee
It follows that the trajectories in figure 1 are described by the equation,
\be
(X^2)^{\mu_1}=c(X^1)^{\mu_2}(m-X^1-X^2)^{\mu_1-\mu_2}
\ee
where the real modulus $c\geq 0$ labels the trajectories and is a measure 
of the separation of the two kinks. The $c=0$ trajectory corresponds to the 
straight-line $3\to 1$ kink, while as $c\rightarrow\infty$, the trajectory 
gets closer and closer to the infinitely separated $3\rightarrow 2\rightarrow 1$ 
trajectories.

\subsection*{$n\geq 3$}

The pattern of kink trajectories described above generalises simply 
to the general case. Consider firstly $n=3$. The vacua now determine 
the points of a right-angle simplex, with the 
solutions \eqn{solution} forming 
its edges. This is shown in figure 2. On each of the four faces of the
simplex, the Bogomol'nyi equations reduce to those of the $n=2$ case 
\eqn{n2bog}, and the trajectories are therefore restricted to lie in the 
face, each of which looks like a copy of figure 1. An analysis of the 
Bogomol'nyi equations near the fixed point at the origin (vacuum 4 in the 
diagram) shows that the trajectories head into the polytope. As each of them 
cannot escape to infinity without crossing the faces, they must end 
on a fixed point. Only those trajectories which lie on the $2-3-4$ face will end at 
vacua $2$ and $3$ and, of those, only those on the $3-4$ edge will end at 
vacuum $3$. All others end at vacuum 1. A typical trajectory is sketched in 
figure 2. We therefore have a two parameter family of kink solutions. These 
parameters have the interpretation of the separation between the $4\rightarrow 3$ kink, 
the $3\rightarrow 2$ kink and the $2\rightarrow 1$ kink. As in the 
$T^\star({\bC\bP}^2)$ 
case, supersymmetry ensures that these separations are paired with angular collective 
coordinates arising from the $\psi_I$.

\begin{figure}
\begin{center}
\epsfxsize=2.0in\leavevmode\epsfbox{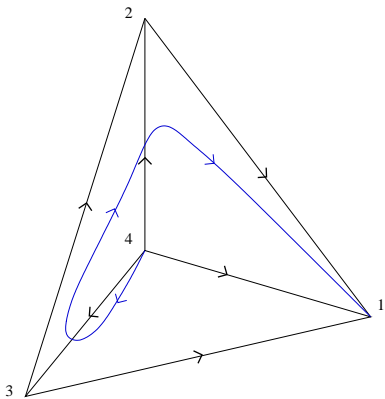}
\end{center}
\caption{\em The BPS flows in the ${\bC\bP}^3$ massive sigma-model. There 
now exists a two-parameter family of kink trajectories corresponding to the 
separation of three kinks. The flows on the faces are copies of figure 1. 
A typical trajectory lying within the tetrahedron is drawn.}
\label{fig2}
\end{figure}

The generalisation of this to $n>3$ is clear. The vacua \eqn{vacua} form the 
vertices of a $n$-dimensional simplex, while the solutions \eqn{solution} 
form the edges. The trajectories on a face of dimension $m$ are determined by 
the Bogomol'nyi equations for $T^\star({\bC\bP}^m)$ and are restricted to lie 
on that face. This bounds the trajectories inside the simplex, each of which 
ends at fixed point 1. Note that in each case the simplex is the toric diagram 
for ${\bC\bP}^n$, with the $n$ angular variables $\psi_I$ providing the 
requisite ${\bf T}^n$ fibre. All trajectories lie within the ${\bC\bP}^n$ 
base of $T^\star({\bC\bP}^n)$ and extend to solutions of the ${\bC\bP}^n$ 
sigma-model itself. 

One may verify that the functions over the simplex,
\be
F(X^I;\alpha_I)= (m-\sum_{I} X^I)^{-\sum_J \alpha_J}\prod_K (X^K)^{\alpha_K}
\ee
are constant on BPS trajectories provided that the parameters $\alpha_I$, 
$(I=1,\cdots,n)$ satisfy, $\sum_I \alpha_I\mu_I=0$. This one constraint on 
$n$ variables ensures that there is an $(n-1)$ parameter family of 
$(n-1)$-dimensional hypersurfaces. For each choice 
of parameters $\alpha_I$, the family of hypersurfaces parametrised by the 
value of $F$ fill the $n$-simplex. Thus, together, $F$ and the $(n-1)$ 
independent $\alpha_I$ yield an $n$ parameter family of $(n-1)$-dimensional 
surfaces. Their intersections are the BPS trajectories.

\section{Dynamics of Domain Walls in $T^\star({\bC\bP}^n)$}

In the previous section, we have seen that the Calabi metric on 
$T^\star({\bC\bP}^n)$ admits a $2n$-dimensional moduli space, ${\cal M}^n$, 
of domain wall solutions interpolating between the first and last vacua. Let 
us denote the collective coordinates  parameterising 
${\cal M}^n$ by $u^a$, $a=1,\cdots,2n$. We have argued that, at least asymptotically, 
these parameters have the interpretation of the position and internal 
degree of freedom of $n$ fundamental kinks. The low-energy dynamics of 
these kinks is given by a sigma-model with four supercharges
on ${\cal M}^n$ with metric given by \eqn{meredith}.  
Given the smoothness of the domain wall solutions, it seems likely that this 
metric is complete. On general grounds, 
we expect the metric to factorise into a centre of mass 
piece, parameterising the overall position of the kinks, together with the 
internal degree of freedom arising from shifts of the tri-holomorphic Killing 
vector field $\mu_Ik^I$. The moduli space is thus
\be
{\cal M}^n={\bR}\times\frac{{\bR}\times\tilde{\cal M}^n}{G}
\label{mod}\ee
where $G$ is a discrete normal subgroup of the isometries. 
Supersymmetry requirements 
ensure that the metric \eqn{meredith} is K\"ahler. Moreover, the symmetries 
of the original massive sigma-model descend to the low-energy dynamics, ensuring 
that the metric on ${\cal M}^n$ is toric K\"ahler i.e. admits $n$ holomorphic $U(1)$ 
isometries\footnote{Note that the potential \eqn{pot} breaks the $SU(n+1)$ 
isometry of the target space to $U(1)^n$ and thus the domain wall moduli 
space inherits only these abelian isometries.}. We denote these by $l^I$, 
$I=1,\cdots,n$. 

There exists a generalisation of the static domain walls that we have been
considering so far to dyonic domain walls, 
or Q-kinks \cite{AT2}. These objects, which 
are $1/2$-BPS \cite{intersect}, carry both topological charge as well as 
Noether charge associated with the isometry $\mu_Ik^I$. They solve the 
time dependent BPS equations \eqn{tink}.
Within the low-energy description of motion on the kink moduli space, they are described by  
excitations along the ${\bR}$ factor of the numerator in \eqn{mod}. 

We would now like to demonstrate the existence of $1/4$-BPS 
Q-kinks and explain how they arise in the low-energy 
dynamics. The analysis is identical to that of $1/4$-BPS monopoles, so we 
will be brief. These objects are related to the intersecting domain wall 
solutions discussed in \cite{intersect}. As in that reference, the 
important point is that the potential \eqn{pot} is not the most 
general potential allowed by supersymmetry. Rather, a HK sigma-model with 
$8$ supercharges 
in $D$ space-time dimensions admits the sum of $(6-D)$ potentials, each 
the length squared of a mutually commuting tri-holomorphic Killing vector 
\cite{bak,intersect}. In order to build $1/4$-BPS objects, we require 
two such potentials and must therefore be in a space-time dimension $D\leq 4$, 
with a target space of dimension $\geq 8$. For the Calabi metrics, we take the 
potential to be of the form,
\be
V=\ft12\mu_I\mu_J(U^{-1})^{IJ}+\ft12\lambda_I\lambda_J(U^{-1})^{IJ}
\ee
The Bogomol'nyi equations for the $1/4$-BPS Q-kinks are derived thus:
\bea
{\cal E}&=&\int {\rm d}x\ \left\{ U_{IJ}({\bf X}^{I\prime}-\mu_K(U^{-1})^{IK}{\bf n})
\cdot({\bf X}^{J\prime}-\mu_L(U^{-1})^{JL}{\bf n}) + U_{IJ}\dot{\bf X}^I\cdot\dot{\bf X}^J
\right. \nn
&& \left.\ \ \ \ \ \  \ \ \ \ \ \ \ \ + (U^{-1})^{IJ}
(\psi^\prime_I+{\bomega}_{IK}\cdot{\bf X}^{K\prime})
(\psi^\prime_J+{\bomega}_{JL}\cdot{\bf X}^{L\prime}) \right. \nn
&& \left.\ \ \ \ \ \  \ \ \ \ \ \ \ \ + (U^{-1})^{IJ}
(\dot{\psi}_I+{\bomega}_{IK}\cdot\dot{\bf X}^K-\lambda_I)
(\dot{\psi}_J+{\bomega}_{JL}\cdot\dot{\bf X}^L-\lambda_J) \right\}\nn
&& +\mu_I[{\bf X}^I\cdot{\bf n}]^{+\infty}_{-\infty} +
\int{\rm d}x\ \left\{\lambda_I(U^{-1})^{IJ}(\dot{\psi}_J
+{\bomega}_{JK}\cdot\dot{\bf X}^K)\right\} 
\eea
The Bogomol'nyi equations are now given by equations \eqn{boggy} and \eqn{boggy2}, 
together with
\bea
\dot{\bf X}^I&=&0 \\
\dot{\psi}_I&=&\lambda_I
\eea
in which case the mass of the Q-kink interpolating between vacua  $I$ and $J$ 
is given by
\be
E_{IJ}=m|\nu_I-\nu_J|+\lambda_KQ^K
\label{energy}\ee
where $\nu_I$ are defined in equation \eqn{sue} and 
\be
Q^K=\int{\rm d}x\ (U^{-1})^{KL}\dot{\psi}_L
\ee
is recognised as the Noether charge associated with the Killing vector field 
$\partial_{\psi_K}$.

As is usual for $1/4$-BPS states, it is possible to rewrite the Noether charge in 
terms of a potential on the kink moduli space ${\cal M}^n$ associated with the 
Killing vectors $l^I$ \cite{me}:
\be
\lambda_IQ^I=G_{ab}(\lambda_Il^{Ia})(\lambda_Jl^{Jb})
\label{chrg}\ee
Finally, note that there is a single condition relating the topological and 
Noether charges which ensures that 
the dyonic state is truely bound rather than, as appears from the energy \eqn{energy}, 
marginally bound. This relation is,
\bea
\lambda_I[{\bf X}^I\cdot{\bf n}]^{+\infty}_{-\infty}
=\int{\rm d}x\ \lambda_I\mu_J(U^{-1})^{IJ} =\mu_IQ^I
\eea
The dynamics of $1/4$-BPS monopoles has been discussed in 
\cite{BLLYone,BLLYtwo,bak,jeromeone,jerometwo}, 
and for instantons in \cite{inst}. In both cases, the low-energy 
dynamics is described by a massive sigma-model on the soliton moduli space. 
The same is true here. The relevant potential on ${\cal M}^n$ is 
given by $V=\lambda_IQ^I$ which, by equation \eqn{chrg}, we know can be 
expressed as the length of a holomorphic Killing vector, ensuring that 
the 4 supercharges of the low-energy dynamics are preserved\footnote{Note 
that the original HK sigma model with two potentials exists in $D\leq 4$ 
space-time dimensions. The kink solutions then have world-volume of dimension 
$d\leq 3$ as is required to construct a massive supersymmetric sigma-model 
description of the dynamics with four supercharges
with a potential given by the length 
of a holomorphic Killing vector.}. The 1/4-BPS 
Q-kinks are then recovered as $1/2$-BPS solutions of the low-energy dynamics. 

There is another class of $1/4$-BPS solitons which may exist in these models, namely 
orthogonally intersecting domain walls. These were discussed in \cite{intersect}. 
In the context of the low-energy dynamics, they occur if the potential $V$ on 
${\cal M}^n$ has more than one isolated minima. In this case, if the original 
domain wall had spatial world-volume dimension $\geq 2$, we could consider a 
``kink-within-a-kink'', in which we build domain wall within the low-energy effective 
theory. We do not know at present if the Calabi metrics admit such 
intersections, but the above observation reduces this question to understanding 
the ${\bf T}^n$ action on ${\cal M}^n$.

\section{Discussion}

We have shown that massive K\"ahler sigma models with compact target 
spaces ${\cal N}$ and massive hyper-K\"ahler sigma models with 
compact target spaces $T^\star({\cal N})$
admit a moduli space of domain wall solutions. We examined these 
solutions in detail for ${\cal N}={\bC\bP}^n$ and showed that 
the collective coordinates of the solution have the interpretation of 
the positions, together with internal degrees of freedom, of $n$ parallel 
fundamental domain walls. The domain wall moduli space is 
identified with the space of Morse flows on ${\cal N}$, where the morse 
function is related to the sigma-model potential on ${\cal N}$. 
There is a natural K\"ahler metric on this moduli space.
 
We close with a few applications. Firstly, there is a 
remarkable similarity between kinks in the ${\bC\bP}^{n}$ model 
and monopoles in $SU(n+1)$ Yang-Mills-Higgs theory. This fact has been noted before 
\cite{AT2,nickone,nicktwo} and is underscored in the present work. Specifically, the 
moduli space of a $(1,1,\cdots,1)$ monopole ($n$ 1's) has a toric 
HK $4n$-metric \cite{connell,gl,lwyone,lwytwo}
. Here we have shown that the moduli space of the highest 
kink in ${\bC\bP}^{n}$ has a toric K\"ahler moduli space of dimension $2n$. 
Moreover, the construction of $1/4$-BPS dyon solutions in the two theories 
is entirely analogous. 
In \cite{nickone,nicktwo} this correspondence between kinks and monopoles was 
made quantitative; it was shown that the BPS mass spectrum of the two 
dimensional ${\cal N}=(2,2)$ ${\bC\bP}^n$ massive sigma-model and the 
four dimensional ${\cal N}=2$ $SU(n+1)$ Yang-Mills theory coincide. This 
correspondence exists at both classical and quantum level. Subsequently, 
it has been understood that the four-dimensional theory also contains 
``$1/4$-BPS-like'' states \cite{jeromeone,jerometwo}\footnote{These states have non-parallel
electric and magnetic charge vectors which 
would make them $1/4$-BPS in the ${\cal N}=4$ theory. However, in the ${\cal N}=2$ 
theory, where no $1/4$-BPS particle states exist, they preserve $1/2$ supersymmetry}. 
The discussion of section 4 suggests that analogous states also exist 
within the two-dimensional ${\bC\bP}^n$ sigma-model. It would be interesting 
to verify this by semi-classical methods. Note that if they do exist, the 
calculation of the central charges performed in \cite{nickone,nicktwo} guarantees that 
their mass coincides with that of the monopoles. 

\begin{figure}
\begin{center}
\epsfxsize=3.0in\leavevmode\epsfbox{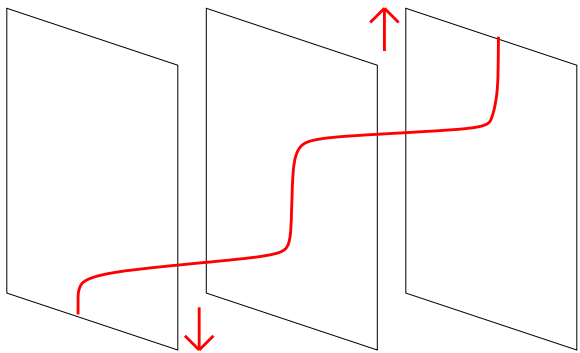}
\end{center}
\caption{\em Twice as kinky: the $D1-D5$ system in background NS $B$-field. The 
single D-string has a modulus in which the two kinks move apart as 
shown by the arrows.}
\label{fig3}
\end{figure}

Finally, we mention an application of our results to the $D1-D5$ system 
of IIB string theory. Consider a single $D1$-brane in the presence
of $n$ parallel but separated $D5$-branes. Turning on a background 
NS $B$-field results in an attractive force between the D-string and 
the $D5$-branes. The D-string has $n$ vacuum states in which it lies within a 
single $D5$-brane, where it appears as an instanton in non-commutative 
$U(n)$ gauge theory, broken to the Cartan subalgebra \cite{nek}.
For small separations between the $D5$-branes, the low-energy dynamics of the 
D-string is described by the massive sigma-model on $T^\star({\bC\bP}^n)$ 
considered in section 3. The kink solutions \eqn{solution} have the interpretation 
of the D-string interpolating from one $D5$-brane to another \cite{paul,lt}. The 
results of section 3 make it clear that this kinky string has a moduli in 
which the two kinks move apart. This is shown in figure 3. It would be interesting 
if one could understand this motion in terms of a gauge theory along the lines 
of \cite{hw}.

\vspace{.5truecm}

\noindent {\bf Acknowledgements}
We would like to thank Bobby Acharya, Fay Dowker,
Ian Dowker, Gary Gibbons, 
Brian Greene, Jae-Suk Park, Sumati Surya, Stefan Vandoren and 
Erick Weinberg for useful discussions. DT would like to thank 
the Korean Institute for Advanced Study and especially Kimyeong Lee 
and Piljin Yi for their hospitality and for several valuable 
discussions during the early stages of this work. DT is also 
grateful to the Center for Theoretical Physics, MIT, for hospitality. 
PKT thanks the Centre Emil Borel of the Institut Henri Poincar{\'e} 
for hospitality. JPG thanks the EPSRC for partial support.
JPG and PKT are supported in part by  PPARC through SPG $\#$613. 
DT is supported by the US Department of Energy.

\end{document}